\documentclass[a4wide,10pt,notitlepage]{article}

\usepackage[T1]{fontenc}
\usepackage[utf8]{inputenc}
\usepackage{lmodern}

\usepackage{amsmath}
\usepackage{graphicx}

\begin{document}







\begin{center}
\textbf{\LARGE Community control in cellular protein production: consequences for amino acid starvation.}

\bigskip

Frank S Heldt$^1$, Chris A Brackley$^2$, Celso Grebogi$^3$ and Marco Thiel$^{3,*}$

\smallskip

\textit{\footnotesize 1. Otto-von-Guericke University Magdeburg, Universitaetsplatz 2, 39106 Magdeburg, Germany.\\
2. SUPA, School of Physics and Astronomy, University of 
Edinburgh, Mayfield Road, Edinburgh, EH9 3JZ, UK.\\
3. Institute for Complex Systems and Mathematical Biology, SUPA, King's College, University of Aberdeen, Aberdeen, AB24 3UE, United Kingdom.}
\end{center}
\noindent{\footnotesize * corresponding author. Email m.thiel@abdn.ac.uk}

\begin{abstract}
Deprivation of essential nutrients can have stark consequences for many processes in a cell. We consider amino acid starvation, which can result in bottlenecks in mRNA translation when ribosomes stall due to lack of resources, i.e. tRNAs charged with the missing amino acid. Recent experiments also show less obvious effects such as increased charging of other (non-starved) tRNA species and selective charging of isoaccepting tRNAs. We present a mechanism which accounts for these observations, and shows that production of some proteins can actually increase under starvation. One might assume that such responses could only be a result of sophisticated control pathways, but here we show that these effects can occur naturally due to changes in the supply and demand for different resources, and that control can be accomplished through selective use of rare codons. We develop a model for translation which includes the dynamics of the charging and use of aa-tRNAs, explicitly taking into account the effect of specific codon sequences. This constitutes a new control mechanism in gene regulation which emerges at the community level, i.e., via resources used by all ribosomes.\\
\textbf{Subject Areas:} biophysics, biomathematics, mathematical modelling

\smallskip

\noindent\textit{A revised version of this manuscript has been published as follows:}\\
\noindent Frank S. Heldt, Chris A. Brackley, Celso Grebogi, Marco Thiel (2015) Community control in cellular protein production: consequences for amino acid starvation \textit{Phil. Trans. R. Soc. A}  \textbf{373} 20150107; DOI: 10.1098/rsta.2015.0107

\end{abstract}


\section{Introduction}

Proteins form the basis of all biological processes within a cell, and one of the key characteristics of life is the ability of an organism to synthesize its own proteins. Understanding how a cell regulates this synthesis remains one of the key questions of modern science. In this paper we propose a new mechanism for protein regulation which originates not from a complex network of gene control pathways, but instead emerges naturally from the interplay between different processes via the supply and demand of different resources. This represents a new paradigm for gene regulation: that of community based control of a complex system.

The amino acids which make up protein molecules are either taken up directly, or formed from precursors found in the cell's environment~\cite{Alberts2002}. When the environment lacks the required nutrients this will have a severe impact on the ability to produce certain proteins; if the cell is to survive, it must react to stresses such as amino acid starvation. As one would expect, if a cell is placed in an environment starved of a particular amino acid, this leads to bottlenecks in the protein production process due to a reduction of the number of amino acid carrying complexes within the cell. In this paper we use mathematical modelling to investigate this scenario. Besides the expected effects, we also uncover some counter-intuitive behaviour which provides insight to the origins of some recent experimental findings~\cite{Dittmar2005f,Zaborske2009}, some of which have previously been unexplained. The process whereby amino acids are assembled into proteins is called mRNA translation, and we consider this in the model organism {\textit{Saccharomyces cerevisiae} (baker's yeast). While previous studies have examined how starvation for amino acids activates feedback mechanisms in the initiation stage of translation~\cite{Hinnebusch2005}, here we consider for the first time the effect on elongation and particularly ribosome traffic. 

The sequence of amino acids which gives a protein its unique structure and function is transcribed from the cell's DNA into mRNA molecules. These are single strands of nucleotides which, in general, encode a single protein sequence; every three nucleotides - one codon - represents one amino acid. Molecular machines called ribosomes translate this information sequentially, moving from codon to codon along the mRNA, adding amino acids to a growing chain. The process begins when a ribosome assembles at the 5' end of an mRNA and scans for the initiation signal (usually an AUG codon)~\cite{Sonenberg2003,Nakamoto2009}. Ribosomes decode each codon via the binding of an aminoacylated-transfer RNA (aa-tRNA)~\cite{Ibba2004}. On encountering the required aa-tRNA the delivered amino acid is incorporated into the chain, and the ribosome moves one codon forward; a tRNA which is no longer bound to an amino acid is released. This process is known as elongation of the amino acid chain. Different codon species are decoded by different tRNA species, each of which can only carry a specific amino acid. We refer to the aa-tRNA complex as a ``charged'' tRNA. Uncharged tRNAs are loaded with their corresponding amino acid by aa-tRNA synthetases~\cite{Delarue1993,Ibba2000}. Upon reaching a stop codon, ribosomes release the complete protein and dissociate from the mRNA. 

There are two factors which determine the rate of protein translation: regulation at the initiation level~\cite{Hinnebusch2005}, and the rate at which ribosomes move during elongation, i.e., the translation rate of each codon~\cite{Soerensen1989}. The latter depends on the time that a ribosome has to wait before finding the correct charged tRNA, and therefore is proportional to the abundance of the aa-tRNAs in the cell~\cite{Varenne1984}. Different tRNA species vary substantially in their abundances~\cite{Ikemura1982,Ikemura1985b}, which gives rise to slow and fast codons. Also, since in yeast there are only 20 amino acids but 41 tRNA species, there is more than one tRNA for some amino acids (isoacceptors). In some cases a slow codon is used when a fast codon for the same amino acid is available. One could reason that the biased use of such codons could regulate the dynamics of translation~\cite{Romano2009}. 

There has been much recent experimental work investigating the effect of amino acid starvation on translation. For instance Dittmar \textit{et al.}~\cite{Dittmar2005f} examined the charging of different isoacceptors of the amino acid subject to starvation. Also, Zaborske {\textit{et al.}~\cite{Zaborske2009} showed that starvation for histidine, leucine or tryptophan causes a decrease in the charging level of not only the tRNA species carrying these amino acids, but also in tRNAs which deliver other amino acids (whose intracellular levels remained unchanged). Intriguingly, some tRNAs even show an \textbf{increased} charging level when cells encounter starvation; \textit{as one of the main results of this paper we provide a mechanism for this counter-intuitive experimental finding}. 

Most previous models of translation have assumed that the abundances of the aa-tRNAs are constant~\cite{MacDonald1968,Shaw2003,Basu2007}. A recent study \cite{Brackley2010,Brackley2010b} included the finite rate of recharging of tRNAs in a probabilistic model of translation, and allowed aa-tRNA abundances to vary dynamically. It has also been shown that the balance between the supply and the demand for tRNA resources is a crucial factor in determining protein production rates~\cite{Brackley2011}. Here we examine the effect of starvation of amino acids on tRNA recharging, and ultimately protein production rates.

One might assume that the interesting effects described above could only occur as a result of complicated feedback mechanisms of control. In this paper we show that actually such behaviour can arise naturally through careful selection of codon sequence. We first use artificial mRNA sequences to show how behaviour such as an increased charging of tRNAs as a response to starvation can be realized, before showing that the same effects occur when we consider mRNA sequences taken from the \textit{S. cerevisiae} genome. We see that regulation of protein production can be achieved through changes in the supply and demand of resources \textit{across all of the mRNAs in the system}, i.e., at the community level. It is possible to control how the protein production from a specific mRNA species responds to changes in the community via a careful selection of coding sequence. We thus propose that there may exist a direct and rapidly responding mechanism of protein regulation at the translational level.

\section{Model}

\begin{figure}
\centering
\includegraphics[width=0.7\textwidth]{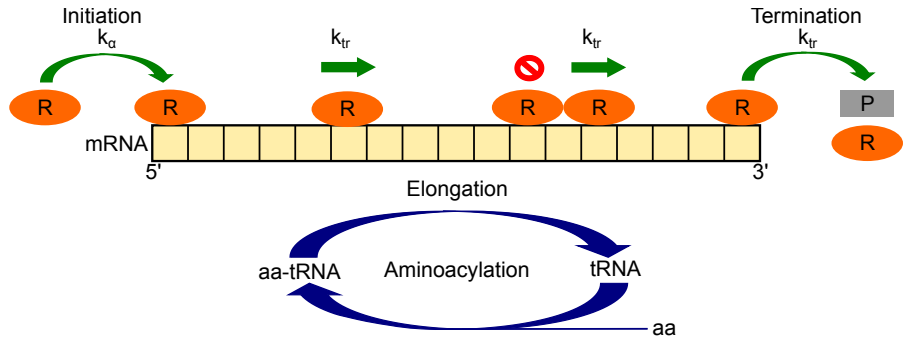}
\caption{\label{fig1} Schematic representation of mRNA translation. An mRNA consists of a sequence of codons, represented here by boxes. Ribosomes ($R$) initiate translation from the 5' end with rate $k_\alpha$. By using aa-tRNAs, they elongate with rate $k_{\rm tr}$, provided the next codon is vacant. When a ribosome reaches the 3' end, it dissociates from the mRNA and releases a protein ($P$). Used tRNAs are recharged by aminoacylation which requires amino acids ($aa$).}
\end{figure}

Figure \ref{fig1} illustrates the two main features of our translation model, namely (i) ribosome traffic along mRNAs and (ii) the cycle of aminoacylation and usage of tRNAs. We also include explicitly the fact that ribosomes extend over several codons and cannot overlap. We now describe each of the aspects of the model in turn.

\subsection{Ribosome traffic}

We derive a system of reaction rate equations for ribosome traffic on mRNAs by adapting a model of translation in bacteria~\cite{Kremling2007}. We consider explicitly that mRNAs are precise sequences of different codons. Ribosomes bind to the mRNA with the rate $k_\alpha$, provided the initiation site and the subsequent codons are vacant
\begin{equation}\label{initiation}
R+\sum_{h=1}^{m} X_h^{\circ} \xrightarrow{k_\alpha} X_1^{\bullet} + \sum_{h=2}^{m} X_h^{\circ},
\end{equation}
where $X_j^\circ$ (or $X_j^\bullet$) represents the $j$th codon of an mRNA. A filled circle denotes a codon occupied by a ribosome's recognition (A) site (which we assume is the leftmost codon covered by that ribosome - note that this choice does not influence the results); an open circle indicates that either the codon is unoccupied, or that it is occupied by a region of the ribosome other than the A site. $R$ represents a free ribosome and $m$ is the number of codons occluded by a translating ribosome. The rate $k_\alpha$ encompasses all of the steps of translation initiation, including the binding of several initiation factors (some of which are themselves regulated under cell stress conditions). Thus $X_1^\bullet$ denotes a ribosome decoding the first codon downstream of the initial AUG. Ribosomes advance in a stepwise manner with rate $k_{\rm tr}$ from codon $j$ to $j + 1$, provided the downstream sites are not occupied. This process requires a charged tRNA that matches the current codon ($T_{C_j}^\star$ , where $C_j$ indexes the species of codon $j$, and the star denotes that the tRNA is charged) and results in the release of its uncharged counterpart ($T_{C_j}$)
\begin{equation}\label{elongation}
X_j^\bullet + T_{C_j}^{\star} + \sum_{h=j+1}^{j+m} X_h^\circ \xrightarrow{k_{\rm tr}}
X_j^\circ + T_{C_j} + X_{j+1}^{\bullet} + \sum_{h=j+2}^{j+m} X_h^{\circ},
\end{equation}
with $j=2,\dots, n$, where $X_j^\bullet$ is a ribosome decoding codon $j$, and $n$ is the mRNA's length in codons. We assume that ribosomes reaching the ($n-m$)th codon continue translation without any hindrance from downstream ribosomes, i.e., we truncate the sums at $h = n$. The translation rate $k_{\rm tr}$ is a constant representing all intermediate reaction steps that the ribosome undergoes~\cite{Basu2007,Fluitt2007}. We approximate this value from the average codon translation rate \cite{Gilchrist2006} and the abundance of tRNAs in yeast estimated from the gene copy number~\cite{Percudani1997}. Note that unlike in other models~\cite{Lakatos2003,Kremling2007}, the overall translation rate of a given codon depends on the concentration of the respective aa-tRNA and hence can vary with time. We make the approximation that the bare tRNA is released instantly after the charged tRNA is captured; in actuality there are several binding sites on the ribosome, and the tRNA remains bound for several further elongation steps. If we assume that it is the availability of amino acids and not bare tRNAs which will limit elongation, then this is unlikely to have any substantial effect on our results.

Eventually, ribosomes reach the stop codon and release a mature amino acid chain ($P$)
\begin{equation}\label{termination}
X_n^\bullet + T_{C_n}^{\star} \xrightarrow{k_{\rm tr}} R + P + T_{C_n} + X_n^\circ.
\end{equation}
The binding of release factors is known to be a fast process~\cite{Heinrich1980,Arava2003}, so we assume that translation of the last codon dominates the rate of termination. Since we do not consider post translational modification of the chain the rate of termination is equivalent to the rate of protein production, which we denote $\lambda$. We describe this system of reactions using a system of ODEs, as detailed in Sec.~\ref{sec:methods} below.

\subsection{Aminoacylation}

The process of charging a tRNA with its cognate amino acid is facilitated by a family of aminoacyl tRNA synthetases. The precise mechanism is still controversial \cite{Delarue1993,Ibba2000} and different synthetases may work via different mechanisms. Thus we do not model charging in detail, but
rather use a simple description
\begin{equation}\label{acylation}
T_k + aa_k \xrightarrow{v_k}  T_k^\star,
\end{equation}
where $aa_k$ represents an amino acid of type $k$. We choose the rate $v_k$ such that the recharging exhibits the kinetics of a two substrate enzyme-catalysed reaction --- an assumption which has been used by several previous authors~\cite{Chassagnole2001,Elf2001}. Hence, the reaction rate is given by 
\begin{equation}\label{vkeq}
v_k = \frac{v_{\mathrm{max},k} [aa_k] [T_k]}{K_{m,aa_k} K_{m,T_k} 
\left(1 + \frac{[aa_k]}{K_{m,aa_k}} \right)
\left(1 + \frac{[T_k]}{K_{m,T_k}} \right) },
\end{equation}
where $K_m$ and $v_{\mathrm{max},k}$ denote the Michaelis--Menten constants and maximum reaction velocity, respectively; these are specific to each aminoacyl tRNA synthetase. Since each elongation step requires a single tRNA, we replace the concentration of tRNA molecules $[T_k]$ in Eq.~(\ref{vkeq}) with the number of molecules $T_k$ and express $K_{m,T_k}$ in numbers of tRNAs. However we still consider the amino acids in terms of their concentration. We also neglect the change in $[aa_k]$ due to the ongoing recharging, and assume that it is constant in time.

\subsection{Simulation Method}\label{sec:methods}

We use reactions~(\ref{initiation})-(\ref{termination}) to derive a set of ODEs for ribosome traffic. We consider a population of N identical mRNAs described by a single set of differential equations. Defining $y_j$ as the proportion of mRNAs where there is a ribosome translating codon $j$ for $1\leq j\leq n$, and $y_j=0$ for $j>n$, yields the equations
\begin{eqnarray}\label{dydt}
\frac{dy_1}{dt} =&
k_\alpha \left[ \prod_{h=1}^m (1 - y_h )\right] - k_{\rm tr} T_{C_1}^\star y_1\left[\prod_{h=2}^{m+1} (1 - y_h )\right],\\
\frac{dy_j}{dt} =&
k_{\rm tr} T_{C_{j-1}}^\star y_{j-1} \left[ \prod_{h=j}^{j+m-1} (1 - y_h )\right] \nonumber \\
& - k_{\rm tr} T_{C_j}^\star y_j \left[ \prod_{h=j+1}^{j+m} (1 - y_h )\right] ~~~~~
\mbox{for}~j = 2,\dots,n.
\end{eqnarray}
Terms of the form $(1-y_j)$ represent hindrance of a ribosome's movement due to the occupation of downstream codons: a ribosome stalls on codon $j$ if any of the codons $j+1$ to $j+m$ are occupied, or it moves with rate $k_{\rm tr} T_{C_j}$ if $j+1$ to $j+m$ are vacant. Given that translation is fast compared to most regulatory mechanisms, we assume that the number of available ribosomes does not change significantly during translation, and that this does not limit the initiation process; i.e., the ribosome availability is incorporated into the initiation rate $k_\alpha$ . For simulations with multiple mRNA species, we will consider multiple sets of ODEs.

For the aminoacylation process described by Eq.~(\ref{acylation}), the number of charged tRNA molecules of species $k$ is described by
\begin{equation}\label{dTdt}
\frac{dT_k^\star}{dt} =
v_k - k_{\rm tr} T_k^\star 
\sum_{C_k} \left[
y_{C_k} \prod_{h=C_k+1}^{C_k+m} (1 - y_h)
\right],
\end{equation}
where $C_k$ indexes codons that require the tRNA species $k$ for decoding.

Equations~(\ref{dydt})-(\ref{dTdt}) form a model of protein translation that explicitly accounts for tRNA charging and its role during elongation. We solve the equations numerically using the SBPD package for the Systems Biology Toolbox 2 \cite{Schmidt2006} in Matlab (version 7.5.0 R2007b). All of the results which we present represent the steady state, i.e., we disregard data from the first part of each simulation to account for transient effects due to the initial conditions.

\section{Results}
In the remainder of this paper, we consider ribosomes that cover $m = 9$ codons and choose an initiation rate $k_\alpha=0.2~\mbox{s}^{-1}$ (under normal conditions ribosomes are spaced on average 50 codons apart and translate 10 codons~s$^{-1}$~\cite{Arava2003,Gilchrist2006}. Although initiation itself is highly regulated under starvation conditions (for example via the GCN pathway \cite{You2011,Hinnebusch2005} or eIF3 initiation factor~\cite{Sha2009}), our focus here is on downstream regulation, so we keep $k_\alpha$ constant -- see Sec.~\ref{sec:discussion} for further discussion. We examine the effect of changing the concentration of one amino acid species on the protein production rate and the charging levels of different tRNAs. We define the latter as the ratio of the number of charged tRNAs of a particular type to total number of tRNAs of that type, and denote it $G_k$. In order to compare different mRNA and tRNA species we normalize the protein production rate $\lambda$ by dividing by the rate found for non starvation conditions. We also consider the probability $y_j$ of finding a ribosome translating the $j$th codon, referring to this as the ribosome density. 

A complete genome wide investigation using the above framework would present a substantial computational challenge; moreover the data such an undertaking would create would be difficult to interpret. For this reason we restrict our simulations to small subsets of the transcriptome, mostly considering only one mRNA sequence at a time. Before analysing simulations using real mRNA sequences, we first illustrate that the use of slow codons can control translation rates using simple artificial mRNA sequences which contain only two codon species, each coding for a different amino acid: a common species which makes up most of the mRNA, and a rare species of which there is only a single codon. We refer to the latter as ``species A'' codons, decoded by tRNAs of type A carrying amino acids of species A. We choose the number of tRNAs of each species such that their ratio is equal to that of the gene copy numbers of the most and least abundant tRNAs in \textit{S. cerevisiae}~\cite{Percudani1997}. Hence, translation of codon species A is slow compared to the translation of the abundant codons. We investigate the effect of starvation of amino acid species A, denoting the concentration $[aa_A]$.

\subsection{Codon sequence determines behaviour at queuing phase transition}

Our results show that the mRNA sequence affects the rate of protein production, and more importantly its response to amino acid starvation. Previous work \cite{Romano2009} has shown that the response of the protein production rate of different mRNAs to variation of the initiation rate depends very much on the position of the slowest codons. This allows for a classification of mRNAs based upon features of the coding sequence; a correlation between the different types of mRNA and the function of the resulting proteins was also found. Inspired by this result, we start by showing that a similar classification arises when we examine the response to starvation. Two different mRNA types emerge in this context: type I mRNAs contain a codon of species A some distance downstream of the 5' region, where a stalling ribosome does not directly impair initiation; type II mRNAs have a codon of species A close to the 5' end, where it is directly covered by initiating ribosomes.

\begin{figure}
\centering
\includegraphics[width=0.7\textwidth]{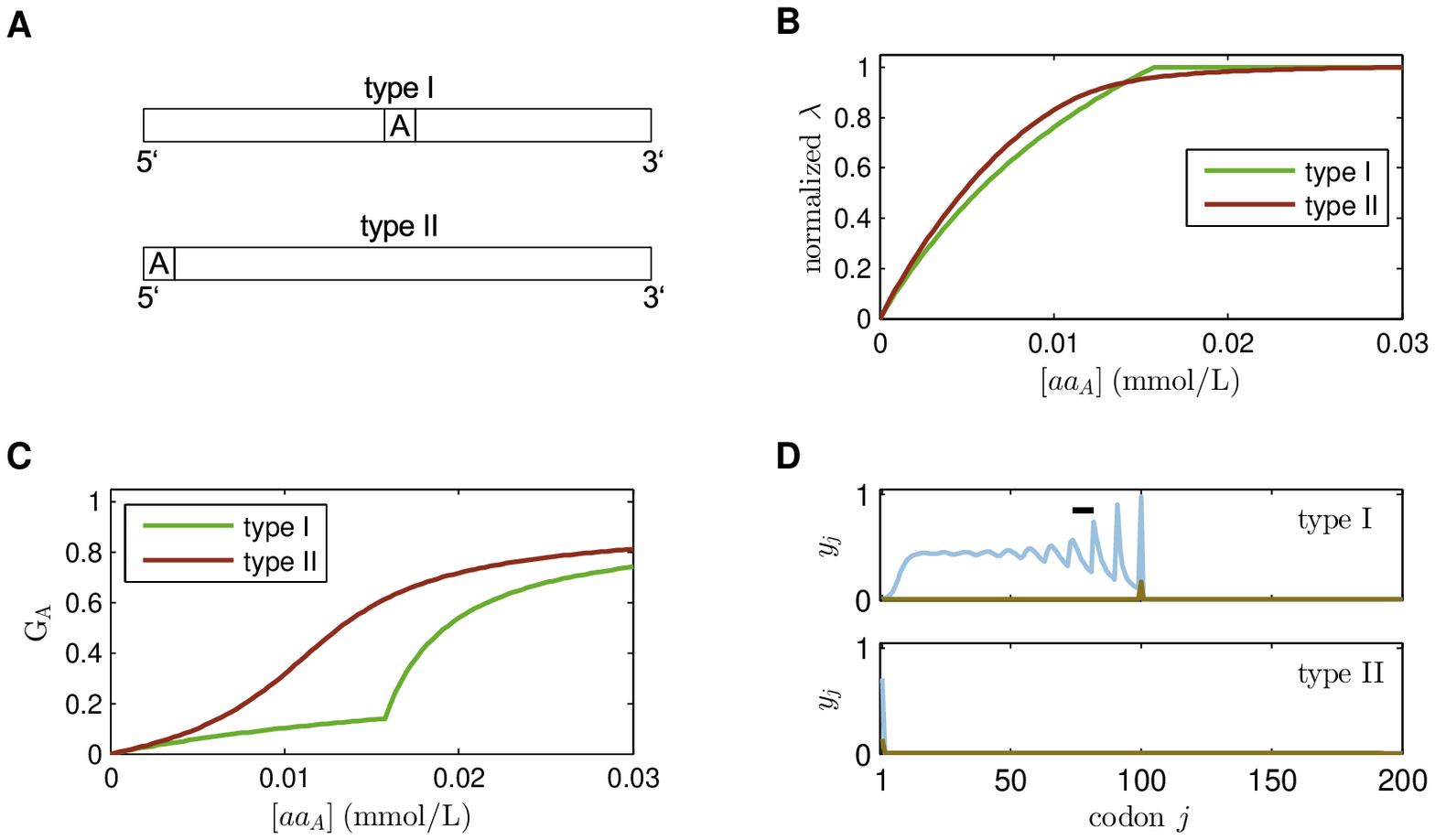}
\caption{\label{fig2} Steady state results from separate treatments of mRNA I and mRNA II: (a) Schematic representation of mRNA types I and II. The position of the codon of species A (corresponding to the starved amino acid) is indicated. (b) Protein production rate per mRNA $\lambda$ (normalized by dividing by the value found for large $[aa_A]$) as a function of the amino acid concentration. (c) Charging level $G_A$ of tRNAs of type A. (d) Codon occupation density at high (brown line) and low (blue line) amino acid concentrations. The oscillatory features in these plots are due to the width of the ribosome, which is indicated by the length of the bar above the line in the upper plot.}
\end{figure}

We illustrate the different behaviour of each mRNA type by treating them in isolation. In each case we consider mRNAs of length $n = 200$; the configurations are shown in figure~\ref{fig2}(a). For simplicity we assume that the recharging parameters for each tRNA species are the same and use experimentally measured values for a common aminoacyl tRNA synthetase: the Leucine-tRNA ligase (EC 6.1.1.4)~\cite{Chirikjian1973,Nawaz2007} . Since the model only considers a subset of the transcriptome, parameters and tRNA numbers are scaled in order to match the demand--supply ratio for tRNAs in a whole cell.

In figure~\ref{fig2}(b), we plot the normalized protein production rate $\lambda$ as a function of $[aa_A]$. We assume that this is constant in time; i.e., we solve for different, constant values of $[aa_A]$. We see a different response for each mRNA.

For type I mRNAs, at large amino acid concentrations, $\lambda$ remains constant as $[aa_A]$ is reduced. Then when a critical concentration is reached there is a sharp change to a regime where $\lambda$ reduces as $[aa_A]$ is reduced (a first--order phase transition). In a model without amino acid starvation a first--order phase transition has been reported to result due to queue formation when the initiation rate is increased through a critical value~\cite{Romano2009}. We show here that a similar effect is seen due to starvation. The lack of amino acid molecules affects protein production by reducing the tRNA charging level [figure~\ref{fig2}(c)] which reduces the translation rate of species A codons. As $[aa_A]$ is reduced translation through this bottleneck becomes the limiting process: a ribosome queue forms. Queue formation upstream of codon A therefore characterizes the translation limited regime [figure~\ref{fig2}(d)]. It is only in this queuing regime (small $[aa_A]$) that $\lambda$ is a function of $[aa_A]$; otherwise the protein production rate is proportional to $k_\alpha$. For type II mRNAs, the protein production rate varies smoothly with $[aa_A]$. A decreased charging level again impairs translation of species A codons which leads to an increased occupation density near the 5' end of the mRNA [figure~\ref{fig2}(d)]. The transition is smooth since the slow codon is at the initiation site, so there can be no sharp onset of queuing.

In summary, for type I mRNAs there is a first order phase transition as $[aa_A]$ is reduced: we move from a regime where protein production is limited by the initiation rate $k_\alpha$ to a regime where it is limited by the slow codon. For type II mRNAs the protein production rate is limited by both initiation and the slow codon; i.e., due to the fact that the first codon is slow, initiation of new ribosomes is hindered.

\subsection{Amino acid starvation can lead to increased tRNA charging}

\begin{figure}
\centering
\includegraphics[width=0.7\textwidth]{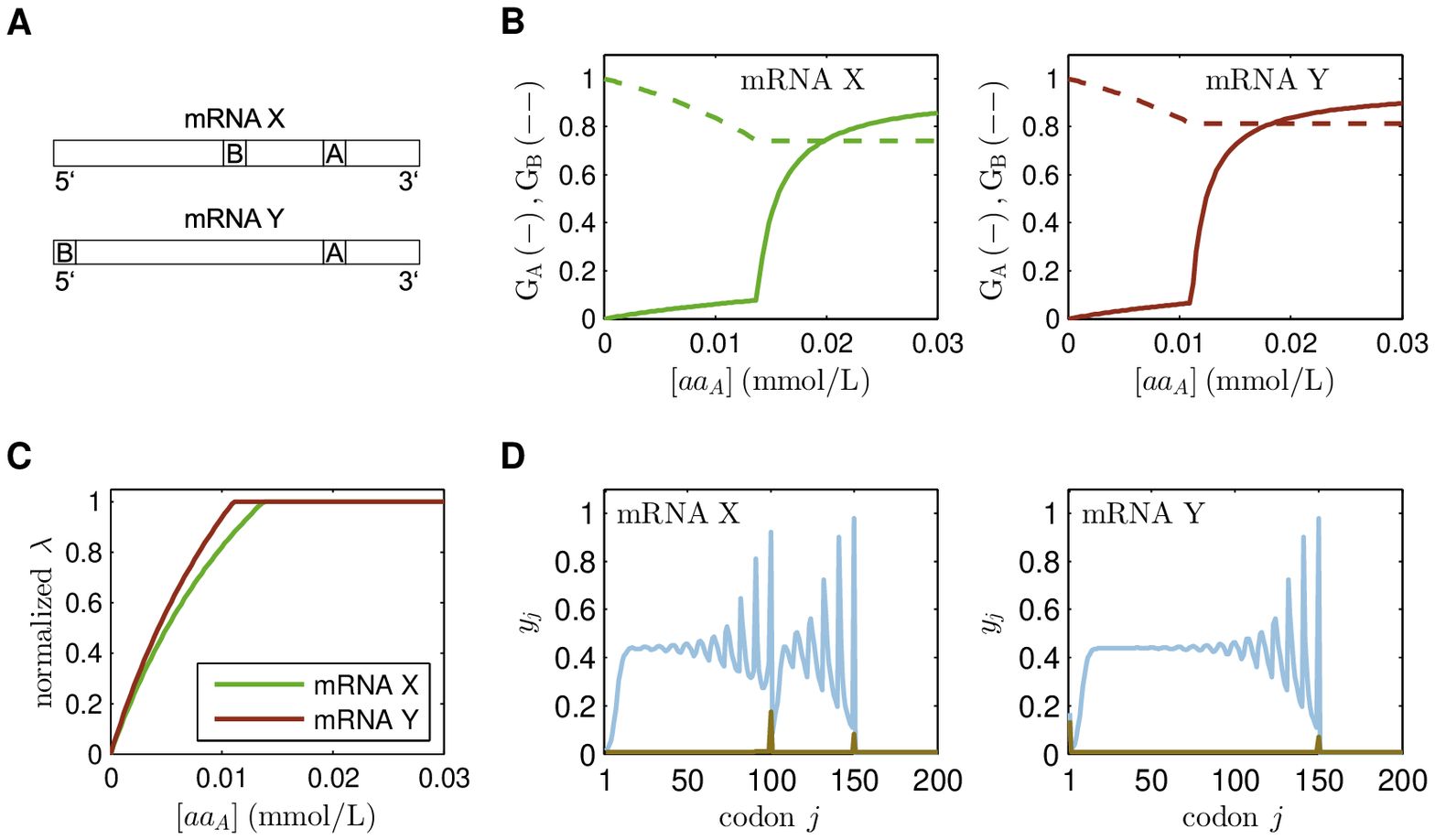}
\caption{\label{fig3}  Steady state results from separate treatments of mRNA X and Y: (a) Schematic representation of both mRNAs. The positions of the codons of species A (abundant tRNA but starved amino acid) and the codons of species B (rare tRNA) are indicated. (b) Charging levels $G_A$ and $G_B$. (c) Normalized protein production rate $\lambda$ as a function of $[aa_A]$.  (d) Codon occupation density at high (brown line) and low (blue line) amino acid concentrations for each mRNA. Although difficult to see in the plot, in the high $[aa_A]$ case for mRNA X there is a small peak in the density at the B-type codon ($j = 100$). }
\end{figure}

Figure~\ref{fig2} demonstrates that amino acid starvation impairs the ribosome traffic. In this subsection we describe how this decreased traffic can affect the charging levels of a tRNA species delivering an amino acid that is not subject to starvation. In figure~\ref{fig3}(a) we introduce two further idealized mRNAs which contain \textit{three} codon species, and we denote these mRNA X and mRNA Y. We again starve for amino acids of species A, but this time these tRNAs are in high abundance. Codons of species B require a rare tRNA which carries an amino acid species that is not subject to starvation. The third codon species makes up the rest of the mRNA and corresponds to an abundant tRNA with constant charging level. 

Under non-starvation conditions species B codons are the slowest, but the initiation rate is not high enough for queuing to occur. The X mRNAs (having their slow B type codon in the bulk) are of type I, and the Y mRNAs (having their slow B codon near the initiation site) are of type II. To clarify our definitions, in non-starvation conditions mRNAs are classified as type I or II depending on the position of the slowest codons.

Upon amino acid starvation, the charging of the A type tRNAs is reduced; for small enough $[aa_A]$, type A codons will become the bottleneck. As in the previous subsection we again treat each mRNA sequence separately, i.e., our system consists of many copies of the same mRNA sequence. From plots of the tRNA charging levels as a function of $[aa_A]$ we observe two distinct phases [figure~\ref{fig3}(b)]. At high amino acid concentrations, variation of $[aa_A]$ only affects the abundance of aa-tRNAs of species A. In this regime, ribosome traffic and protein production remain constant [figure~\ref{fig3}(c)] as the availability of tRNAs corresponding to the slowest codons (species B) is unchanged. When the starved amino acids become very
rare the system enters a second regime where codons of species A become the slowest; i.e., due to starvation charged tRNAs of type A have become depleted, and this causes ribosomes to queue [figure~\ref{fig3}(d)]. Since for both mRNAs the species A codon is in the bulk, in this regime both mRNAs are effectively of type I. That is to say, although under non-starvation conditions mRNA Y is of type II, under species A amino acid starvation it shows a type I response - the codon which becomes slowest is far from the initiation site.

Intriguingly, for both mRNAs in the low $[aa_A]$ regime the charging level of (the non-starved) type B tRNA increases with decreasing amino acid concentration. Increased tRNA charging in response to amino acid starvation has been observed experimentally~\cite{Zaborske2009}, and our model now provides a theoretical explanation for this phenomenon (see below). Decreasing $[aa_A]$ causes queues to form upstream of species A codons, reducing the overall elongation rate. This reduces the rate at which species B codons are used by the ribosomes, resulting in an increased charging level of B type tRNAs.

The difference between mRNA X and Y is that when a slow site is near the initiation site it effectively reduces the initiation rate. Since queuing depends on the ratio of the initiation rate and the translation rate of the slowest codon (under starvation this is species A), for mRNA Y $[aa_A]$ must reach a lower value before we see queuing [figure~\ref{fig3}(c)].

\subsection{Amino acid starvation can result in increased protein production}

The above results, along with those of previous studies~\cite{Robinson1984,Soerensen1989,Romano2009}, indicate that specific codon usage plays a significant role in determining the translational dynamics of an mRNA. However, the transcriptome is comprised of thousands of different mRNAs~\cite{Hereford1977a}, and ribosomes translating these mRNAs all use a common pool of aa-tRNA resources. In particular, when resources are sparse different codon usage might govern the effectiveness with which ribosomes translating different mRNA species can utilize these resources.

\begin{figure}
\centering
\includegraphics[width=0.7\textwidth]{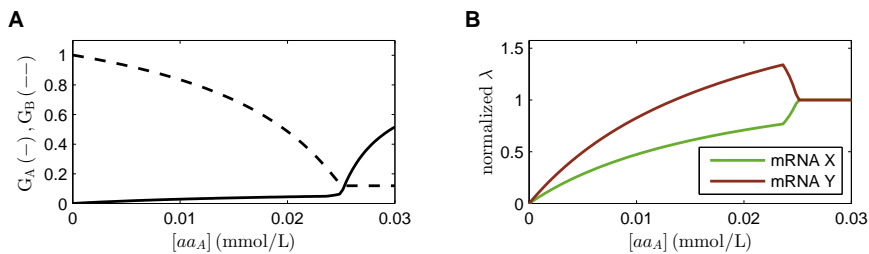}
\caption{\label{fig4} The two mRNA configurations depicted in figure~\ref{fig3}(a) are treated simultaneously and have to compete for a common pool of charged tRNAs. (a) Charging levels $G_A$ and $G_B$, and (b) normalized protein production rate $\lambda$ as a function of $[aa_A]$.}
\end{figure}

We use the two mRNA configurations introduced in figure~\ref{fig3}(a) to explore this scenario in more detail. In contrast to the previous subsection, here both sequences are included in the same simulation, using the same aa-tRNA pool. The abundance of aa-tRNAs [figure~\ref{fig4}(a)] follows similar dynamics as presented for separate translation: a decrease in the species A amino acid concentration results in a decreased charging level of A type tRNAs and an increased charging of B type tRNAs. However the response of the protein production rates from each mRNA species [figure~\ref{fig4}(b)] stands in sharp contrast to that of the previous subsection [figure~\ref{fig3}(c)]. \textit{There is now a regime where there is an increased protein production rate from Y type mRNAs as a result of a decrease in the amino acid concentration.} To our knowledge this is the first time that such an increase has been shown in a model of protein translation, and this is a key result in the present work. In this regime, codons of species A are flow-limiting for the X mRNA; however for the Y mRNA, due to its proximity to the initiation site it is still the codon of species B which is the bottleneck. Since the flow on the X mRNA is reduced, the rate at which ribosomes on this mRNA use type B tRNAs is reduced. Thus ribosomes on Y mRNAs are able to use these type B tRNAs, and the protein production rate increases. As $[aa_A]$ is further reduced the system moves into the familiar regime where protein production from both mRNA species reduces; yet even here Y mRNAs maintain a higher (normalized) $\lambda$ compared with their X counterparts.

In summary we have shown that even with very simple designer mRNAs the codon configuration can significantly affect the protein production performance. Real mRNA sequences consist of codons representing up to 41 tRNA types: they are inherently more complicated. This means that there is even more opportunity for control of protein production given different starvation conditions. For example it might be possible to choose a sequence such that translation of a particular transcript is enhanced in response to starvation of one amino acid species, but is reduced in response to starvation of another.

\subsection{Realistic mRNA sequences}

In order to investigate whether the qualitative behaviour obtained for the mRNA types (I and II) which we identified above also emerge for realistic codon configurations in the context of their response to starvation, in this section we consider the translation of two yeast mRNAs (YLR382C and YHR208W) during leucine starvation. These sequences code for proteins taking part in tRNA charging and amino acid metabolism, respectively: the leucine-tRNA ligase (LeuRS, EC 6.1.1.4), which charges tRNAs with leucine; and the branched-chain-amino-acid transaminase (Transaminase, EC 2.6.1.42), which degrades leucine. Note that we do not consider the function of the proteins in our model but we can speculate on why it might be beneficial to the cell to have a different response for different proteins.

\begin{figure}
\centering
\includegraphics[width=0.7\textwidth]{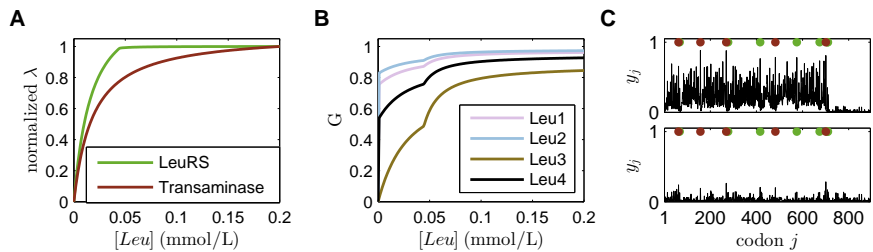}
\caption{\label{fig5} Steady state results from separate treatments of the mRNAs coding for the proteins YLR382C (LeuRS) and YHR208W (Transaminase): (a) Normalized protein production rate $\lambda$ as a function of the leucine concentration. (b) Charging levels of the four leucine-tRNA isoacceptors (Leu1 (UUA), Leu2 (UUG), Leu3 (CUC, CUU) and Leu4 (CUA, CUG)) for the case of LeuRS mRNA translation. (c) Codon occupation density of the LeuRS mRNA at low (upper panel) and high (lower panel) amino acid concentrations. Using dots along the top of the plots, we indicate the positions of the two types of codon which are decoded by the aa-tRNA which becomes most depleted, i.e., Leu3 (CUU in red and CUC in green).}
\end{figure}

Figure~\ref{fig5}(a) shows results for each of the two real mRNA sequences treated separately. We can clearly identify different kinds of behaviour in each case during starvation for leucine. Protein production of the LeuRS mRNA shows a sharp phase transition indicating type I behaviour, whereas translation of the transaminase follows type II characteristics. Hence in the model, a decrease in the leucine concentration directly affects the production rate of the leucine transaminase - we speculate that this is a desired effect, since this will reduce further leucine degradation. Conversely, translation of the mRNA for the leucine recharging enzyme is robust to changes in the leucine concentration down to a much lower level.

In figure~\ref{fig5}(b) we plot the charging levels of the four different leucine tRNAs for the case of LeuRS mRNA translation. Our results resemble previous experimental data \cite{Dittmar2005f} in that the four tRNA species are differently charged. Specifically the rare tRNA species Leu3 and Leu4 become rapidly discharged in response to starvation, whereas the abundant tRNAs (Leu1 and Leu2) maintain a higher charging level. This pattern is determined solely by the concentration of the isoacceptors as we have not included any additional selective recharging in the model.

If we consider ribosome traffic [figure~\ref{fig5}(c)], we observe that the reduced availability of aa-tRNAs again leads to ribosome queue formation. We show on the plot the positions of the two codon species which are decoded by the Leu3 tRNA, the charging level of which becomes most depleted. These are the CUC codon, which matches the Leu3 anticodon exactly, and the CUU codon which due to wobble base pairing is translated even more slowly~\cite{Gilchrist2006}.

\subsection{Community based protein regulation can explain experimental results}

As noted previously, recent experiments on amino acid starvation have uncovered unexpected behaviour which can be explained by the above results. Dittmar et al. \cite{Dittmar2005f} find that different isoaccepting tRNAs for the same (starved) amino acid show different levels of charging. We have shown above that this can occur naturally due to different abundances of the tRNAs and codons, i.e., the balance between supply and demand. This is consistent with the previous prediction of~\cite{Elf2003}.

Zaborske et al. \cite{Zaborske2009} investigated starvation for different amino acids \textit{in vivo} using micro array techniques. They found that the charging level for the staved amino acids decreases, as do charging levels of some other tRNAs (which they show is due to the general amino acid control pathway). Unexpectedly some tRNA species also show an increase in charging level; as we have shown here this could result from decreased rate of use of aa-tRNAs due to bottlenecks forming on some mRNAs. We propose that a method for testing whether this behaviour does occur due to community competition for resources would be an \textit{in vitro} experiment where only the translational machinery is present, i.e., without other control pathways.

\section{Discussion}\label{sec:discussion}

We have developed a model of protein translation which considers variable codon translation rates that depend on the availability of aa-tRNAs. The framework explicitly accounts for aminoacylation and allows us to examine the influence of amino acid starvation on ribosome traffic. It leads to a new method of protein regulation where the use of slow codons determines how translation of an mRNA responds to changes in the availability of resources in the wider community.

In the past, stochastic models have revealed the existence of two distinct mRNA types depending on the position of the slowest codon, as well as a relationship between the mRNA type and the protein function~\cite{Romano2009}. Using simple artificial mRNA sequences, we have found the same two types to yield different responses to amino acid starvation. An mRNA showing a type I response to starvation of a particular amino acid is, in terms of protein production, robust against fluctuations in amino acid concentration provided it remains above a critical level. In contrast, the translation of mRNAs showing a type II response is impaired by any change in amino acid concentration. The type of response of a given mRNA may depend on which amino acid is subject to starvation.

The two principal response types were also identified in realistic yeast mRNAs. We therefore propose that proteins can be categorized with respect to starvation for a specific amino acid: those which are robust to changes in amino acid concentration, and those which are highly affected. We have presented a more natural example where the production of a protein involved in aminoacylation is robust, but the production of a protein responsible for degrading that amino acid responds strongly. Although we cannot draw any firm conclusions from our limited simulations, one could hypothesize that the production of proteins crucial to cell function could be prioritized over that of others using this mechanism. This would be consistent with the correlation between slow codon arrangement and protein function reported in~\cite{Romano2009}, as well as other recent studies which show that many mRNA sequences have slow codons positioned close to the initiation site~\cite{Tuller2010,Cannarozzi2010}.

When considering a codon species whose tRNA delivers an amino acid which is not subject to starvation, we find that the charging level of this species can in fact increase in response to starvation. This agrees with previous studies which suggest that tRNA charging levels depend on the balance between their supply and demand~\cite{Elf2003,Elf2005,Brackley2011}, and the phenomenon has also been observed experimentally~\cite{Zaborske2009}. Reduced protein production from some mRNAs can free up resources, which can then be used by other mRNAs leading to increased protein production.

In contrast to previous models we have explicitly taken the codon sequence into account. Our description is, however, deterministic and is essentially a mean field treatment. This latter point refers to the fact that since $y_i$ refers to an average over mRNAs (see Materials and Methods), we do not take into account any spatial correlations in the density. Although correlations are likely to have a large impact on density profiles around bottleneck sites, by comparing with stochastic models which do take correlations into account \cite{Derrida1992,Brackley2010b} we predict that the overall behaviour is unlikely to change qualitatively. We have also neglected any active variation in the initiation rate which may result as the cell actively responds to starvation: for example, the GCN pathway \cite{You2011,Hinnebusch2005} gives rise to a global reduction in initiation rate, while other pathways act to increase or reduce translation of specific mRNA species. Nevertheless, by keeping $k_\alpha$ constant in this study we have been able to identify a translation control method based on tRNA and codon usage. Whilst taking into account variation in initiation rates (and also variation of the abundance of different mRNA species~\cite{Gingold2011}) would no doubt lead to a more complicated response to starvation, one might expect an even more important role for tRNA charging dynamics as regulation at different levels changes the pattern of supply and demand~\cite{Brackley2011}.

In summary, we find that counter-intuitive effects emerge when considering amino acids as a constrained resource. Starvation can lead to different tRNA species having decreased or increased charging levels depending on the codon configuration of all transcribed mRNAs. This mechanism opens a new perspective on protein production control which emerges naturally at a community level due to the complex nature of the system.

\section*{Acknowledgements}

The authors thank M. C. Romano, I. Stansfield, L. Ciandrini, A. Kort, and M. Rehberg for helpful discussions. This work was funded by BBSRC grants BB/F00513/X1 and BB/G010722, and the Scottish Universities Life Science Alliance (SULSA).




\end{document}